\documentclass[preprint,showpacs,preprintnumbers,amsmath,amssymb,
tightenlines
]{revtex4}

\usepackage{graphicx,color}
\usepackage{subfigure}

\newcommand{\de}[2]{(\delta_{#1}^d)_{#2}}

\begin{document}

\preprint{KIAS--P05019}

\title{Re ($\epsilon^{'} / \epsilon_K$) vs. 
$B_d \rightarrow \phi K_S$ CP asymmetry }


\author{P. Ko}
\email[]{pko@kias.re.kr}
\affiliation{School of Physics, KIAS, 
Cheongnyangni-2-dong, Seoul, 130--722 Korea}

\author{A. Masiero}
\email[]{antonio.masiero@pd.infn.it} 
\affiliation{Dipt.\ di Fisica G.Galilei, Univ.\ di Padova,
and INFN, Sezione di Padova, 
Padova, Italy
}

\author{Jae-hyeon Park}
\email[]{jhpark@kias.re.kr}
\affiliation{School of Physics, KIAS, 
Cheongnyangni-2-dong, Seoul, 130--722 Korea}


\date{March 11, 2005}

\begin{abstract}
In a SUSY GUT seesaw scenario, the largeness of the atmospheric neutrino mixing 
can reflect itself into an enhanced flavor changing mixing of beauty and 
strange right-handed scalar quarks. 
If the CP violating phase in  such down type squark $RR$ insertion is the 
main source of CP asymmetry in $B_d \rightarrow \phi K_S$ and 
the gluino contributions to $K^0$--$\overline{K^0}$ and 
$B^0$--$\overline{B^0}$ mixing are negligible, 
there is a correlation between Re ($\epsilon^{'} / \epsilon_K$) and 
$B_d \rightarrow \phi K_S$ CP asymmetry, in addition to that with the 
strange quark  CEDM. The current data on 
Re ($\epsilon^{'} / \epsilon_K ) = (16.7 \pm 2.6 ) \times 10^{-4}$ 
imply that $S_{\phi K}$ should be greater than $\sim 0.5 ~(0.25) $ for 
$\mu \tan\beta = 1 ~(5)$ TeV, assuming the $RR$ dominance in 
$b\rightarrow s$ transition and the minimal 
supergravity type boundary conditions for soft parameters. 
\end{abstract}


\maketitle


\section{Introduction} 

The flavor changing neutral current (FCNC) processes induced by 
$b \rightarrow s$ transitions have played a
major role in probing proposals of new physics at the electroweak scale. 
While the study of $b \rightarrow s \gamma$ keeps being the highlight in
such investigations, more recently the relevance of $b \rightarrow s$
induced purely hadronic decays has been emphasized, in particular in
relation with the issue of testing CP violation in $B$ physics. Among such
decays, the process $B_d \rightarrow \phi K_S$ has aroused much interest 
\cite{worah} at
least for two reasons: i) the process can occur only at the loop level in
the Standard Model (SM), hence making it particularly suitable to spot
sizeable contributions coming from new physics; ii) in the SM the CP
asymmetry in such decay arises only from the indirect CP violation of the 
$B$ mixing, hence one can safely state that within the SM the golden mode
$B_d \rightarrow J/\psi K_S$ and the decay $B_d \rightarrow \phi K_S$ yield
the same amount of CP asymmetry, namely $S_{\phi K} = S_{\psi K}$; if the
new physics entails the presence of direct CP asymmetry in the
decay amplitude of  $B_d \rightarrow \phi K_S$, this can be revealed by
a departure from the equality between the two mentioned CP
asymmetries \cite{kagan,moroi,lunghi,chang,causse}. Indeed, the first data  
on the CP asymmetry in $B_d \rightarrow \phi K_S$ in 2002 \cite{2002} gave 
rise to a wave of interest on this
potentially very interesting signature of new physics which still 
continues today \cite{Hiller:2002ci,Kane:2002sp,
Harnik:2002vs,Ciuchini:2002uv,Endo:2004xt,Hisano:2003bd,Ciuchini:2003rg,%
Hisano:2004tf,Kane:2004ku}.
 
Indeed, even though the discrepancy of the data with respect to the SM 
predictions has been constantly decreasing in the two years elapsed from 
the first results in 2002, still the current world averages of $S_{\phi 
K}$ and $S_{\psi K}$ are \cite{lp04}
\[
S_{\phi K} = ( 0.34 \pm 0.21 ), ~~~~S_{\psi K} = ( 0.726 \pm 0.037 ),
\]
namely  $S_{\phi K}$  is about 2 $\sigma$ lower than the SM prediction 
$S_{\phi K}^{\rm SM} \simeq S_{\psi K}^{\rm SM}$. 
 
Notice that even if one does not wish to speculate much on the single 
above discrepancy, it is interesting to note that the data concerning 
$b \rightarrow q \bar{q} s$ processes with $q=u,d,s$ exhibit an overall 
discrepancy in the values of the CP asymmetry with respect to the SM 
predictions (indeed, the CP asymmetries in $B\rightarrow \phi K_S$ and 
$B_d \rightarrow \eta^{'} K_S$ turn out to be smaller than that
in $B_d \rightarrow J/\psi K_S$ which is measured from
the tree level process $b \rightarrow c \bar{c} s$).
This discrepancy might be a signal for physics beyond the SM
\cite{yamaguchi,murayama,Khalil:2004wp,Chankowski:2005jh}. 

Assuming that the current low value of $S_{\phi K}$ is a signal of new 
physics, we need a new CP violating phase in $b\rightarrow s$ transition. 
An attractive possibility for such new physics beyond the SM occurs in  
supersymmetric grand unified theories (SUSY GUT's)  scenarios with 
seesaw mechanism for neutrino masses and mixings. In such scenarios, 
the large atmospheric neutrino oscillation can be related with a
large $b\rightarrow s$ transition through down type squark and gluino loop 
effects. This flavor changing effect is parameterized by 
a mixing parameter $( \delta^d_{23} )_{RR}$ with a CP phase $\sim O(1)$. 
We follow  Ref.~\cite{Gabbiani:1996hi} for the definitions of the mass 
insertion parameters $( \delta_{ij}^d )_{AB}$'s. 
Let us note that the following  relations
\[
\de{ij}{LL} = \de{ji}{LL}^* , ~~~\de{ij}{RR} = \de{ji}{RR}^* , 
~~~\de{ij}{LR} = \de{ji}{RL}^* ,
\]
will be useful, when we relate $\de{23}{RR}$ and $\de{32}{RR}$ in this work.
For low $\tan\beta$, the single $RR$ insertion can lead to some deviation
in $S_{\phi K}$, if gluinos and squarks are relatively light.
For large $\tan\beta$ case, the double mass insertion can lead to an effective
$RL$ insertion of $O(10^{-2})$, leading to a significant deviation 
in $S_{\phi K}$ from the SM prediction.
In Fig.~1 (a), we show the Feynman diagram for $b\rightarrow s g$ 
involving a CP violating $( \delta^d_{23} )_{RR}$. 

The new CP violating phase in the $RR$ insertion can affect the strange 
quark chromo-electric dipole moment (CEDM) through triple mass insertions, 
if there is an $LL$ insertion between 
third and second generation down squarks.  
[ Fig.~1 (b) ] \cite{Hisano:2004tf,Kane:2004ku,Abel:2004te}. 
Since the $LL$ insertion is generically present in the minimal 
supergravity (mSUGRA) case,  the strange quark CEDM puts a strong 
constraint on the possible deviation of $S_{\phi K}$ from the SM prediction. 
However, the substantial theoretical uncertainties occurring when one  
relates the quark CEDM's with the hadronic EDM's suggest that 
it would be preferable to have some other observable at disposal 
in addition to the strange quark CEDM in order to constrain $S_{\phi K}$.

In this letter, we point out that the  phase in the 
$( \delta_{23}^d )_{RR}$ mixing parameter that would affect $S_{\phi K}$ 
can also contribute to direct CP violation within the neutral kaon system, 
namely Re ($\epsilon^{'} / \epsilon_K$) through triple mass insertion.
The  Feynman diagram for Re ($\epsilon^{'} / \epsilon_K$) with triple 
mass insertion [ Fig.~1 (c) ] is very similar to the Feynman diagram for 
the strange quark CEDM [ Fig.~1 (b) ].  
Needless to say, making use of the observable Re ($\epsilon^{'} / 
\epsilon_K$) to constrain some SUSY soft breaking parameters also entails 
theoretical uncertainties mainly ascribed to our ignorance in the 
evaluation of the relevant hadronic matrix elements. Our discussion shows 
that, even taking into account such a huge degree of uncertainty, 
Re ($\epsilon^{'} /\epsilon_K$) still constitutes a precious tool in 
constraining the interesting flavor changing  mass insertion parameter 
$( \delta^d_{23} )_{RR}$, and, in 
any case, it plays at least a complementary role to the strange quark CEDM 
in performing such a task.   
  Although the large $RR$ mixing is well motivated in
  SUSY GUT plus seesaw mechanism,
  we should emphasize that the link between
  Re ($\epsilon^{'} /\epsilon_K$) and $S_{\phi K}$
  is a generic feature of $RR$ dominance scenario
  which in general can arise in other contexts.

For definiteness, we will work in the mSUGRA boundary condition at the 
reduced Planck scale $M_* \simeq 2.4 \times 10^{18}$ GeV, namely the flavor
universal scalar mass parameter $m_0$, and the trilinear couplings $A$
which is proportional to the Yukawa couplings. Then flavor changing off 
diagonal squark masses will be induced by the renormalization group (RG)
evolution from $M_*$ down to $M_{\rm GUT} \simeq 2 \times 10^{16}$ GeV,
and subsequently from there to the $M_W$ scale 
\cite{Hagelin:1992tc}: 
\begin{equation}
  ( m^2_{\tilde{q}} )_{ij} = - \frac{1}{8 \pi^2} 
  [V^\dagger \lambda_u^2 V]_{ij} 
  ( 3 m_0^2 + A^2 ) \left( 3 \log \frac{M_*}{M_{\rm GUT}} + 
   \log \frac{M_{\rm GUT}}{M_{W}}      \right),
\end{equation}
from which
one can estimate \cite{Hisano:2003bd,Hisano:2004tf} 
\[
\de{13}{LL} \simeq 8 \times 10^{-3} \times e^{- i 2.7} . 
\]
The $LR$ or $RL$ mixing will be small in this limit, and we ignore
them in the following.

For the $RR$ insertion, it is also generated by the RG evolution. In the 
presence of a seesaw mechanism to give rise to neutrino masses, there 
will be the new Yukawa couplings responsible for the Dirac entries in the 
neutrino mass matrix. In a unified context, such couplings intervene also 
in the evolution of the right-handed squark masses in the interval from 
the scale of appearance of the soft breaking terms down to the scale of 
right-handed neutrino masses. If at least some of such couplings are large 
(for instance one of the Dirac neutrino couplings could be of the order 
of the top Yukawa couplings), then one can expect sizeable contributions 
to the RG induced off-diagonal entries in right-handed squark mass matrix.
Let us work in the basis where the down quark
and the charged lepton mass matrices are diagonal.
In this particular basis, we get \cite{moroi}
\begin{equation}
  \label{eq:m2dij}
  \begin{aligned}
    ( 
  m^2_{\tilde{d}} )_{ij} &\simeq - \frac{1}{8 \pi^2} 
  [Y_N^\dagger Y_N ]_{ij}  ( 3 m_0^2 + A^2 ) \log \frac{M_*}{M_{\rm GUT}}
  \\
  &
  \simeq - \frac{1}{8\pi^2} e^{-i (\phi^{(L)}_i - \phi^{(L)}_j)}
  y_{\nu_k}^2 [V_L^*]_{ki} [V_L]_{kj} ( 3 m_0^2 + A^2 ) 
  \log \frac{M_*}{M_{\rm GUT}} .
  \end{aligned}
\end{equation}
Here $Y_N$ denotes the $3 \times 3$ Yukawa matrix of the term
inducing the neutrino Dirac mass, 
which is diagonalized in the form,
\begin{equation}
  Y_N = U_N^\dagger \widehat{Y}_N V_L \widehat{\Theta}_L , \quad
  \widehat{Y}_N = \mathrm{diag} \left(
    y_{\nu_1}, y_{\nu_2}, y_{\nu_3}
    \right), \quad
  \widehat{\Theta}_L = \mathrm{diag} \left(
    e^{i \phi^{(L)}_1}, e^{i \phi^{(L)}_2}, e^{i \phi^{(L)}_3}
    \right) ,
\end{equation}
where $V_L$ is a unitary matrix with a single phase
in the standard parameterization, $U_N$ is a general unitary matrix,
and the phases are subject to the constraint,
$\phi^{(L)}_1 + \phi^{(L)}_2 + \phi^{(L)}_3 = 0$.
Hereafter, we assume that the right-handed neutrino mass matrix
$M_N$ is diagonal
in the basis where $Y_N$ is diagonal,
\begin{equation}
  M_N = U_N^\dagger \widehat{M}_N U_N^*, \quad
  \widehat{M}_N = \mathrm{diag} \left(
    M_{N_1}, M_{N_2}, M_{N_3}
    \right) .
\end{equation}
The simplest case of this sort would arise for $M_N$
proportional to a unit matrix.
A case with non-degenerate eigenvalues is also possible.
Such a texture with simultaneously diagonalizable $Y_N$ and
$M_N$ may result from simple U(1) family symmetries.
Under this assumption, $V_L$ coincides with 
hermitian conjugate of the usual definition of
PMNS lepton mixing matrix, which is not always the case
for arbitrary $M_N$.

From Eq.~(\ref{eq:m2dij}), we can estimate 
$\de{ij}{RR}$.
In particular, having in mind a large Yukawa coupling of the third 
generation, the entry $( \delta_{23}^d )_{RR}$ would be largely affected:
\begin{equation}
\label{eq:d23RR}
\de{23}{RR} \simeq 2 \times 10^{-2} ~
\left( {M_{N_3} \over 10^{14}~{\rm GeV}} \right).
\end{equation}
Its size depends on the right-handed neutrino mass scale, because
$y_{\nu_3}^2$ in Eq.~(\ref{eq:m2dij}) grows with $M_{N_3}$
for a fixed value of neutrino mass.
Sizes of $\de{12}{RR}$ and $\de{13}{RR}$ crucially depend on
$y_{\nu_1}$, $y_{\nu_2}$, and $[V_L]_{31}$ as well.  In general,
they can be large enough to influence $\epsilon_K$,
Re ($\epsilon^{'} / \epsilon_K$),
and $B^0$--$\overline{B^0}$ mixing significantly.
Suppose that the neutrino mass spectrum has normal hierarchy
with negligible lightest neutrino mass.
In this case, Eq.~(\ref{eq:m2dij}) implies that
$\de{12}{RR}$ and $\de{13}{RR}$ scale as
\begin{equation}
  \label{eq:1213}
    |\de{12}{RR}| \sim
    |\de{13}{RR}| \sim \max \left(y_{\nu_2}^2 / y_{\nu_3}^2 ,
      \left|[V_L]_{31}\right|
    \right) \times |\de{23}{RR}| ,
\end{equation}
where we regard all the elements of $V_L$ as $O(1)$ except for $[V_L]_{31}$.
Maximal size of $\de{23}{RR}$ that can be expected from
Eq.~(\ref{eq:d23RR}) is $O(0.1)$
for neutrino Yukawa couplings to be perturbative.
If the right-handed neutrino masses are degenerate so that
$y_{\nu_2}^2 / y_{\nu_3}^2 \simeq 0.2$,
the resulting $\de{12}{RR}$ can result in a huge contribution to $\epsilon_K$
\cite{Gabbiani:1996hi,Moroi:2000mr}.
Also $|[V_L]_{31}|$ close to the current upper bound $\sim 0.15$
\cite{Maltoni:2004ei} can give rise to a similar size.
For large $\tan \beta$,
this can lead to contribution to Re ($\epsilon^{'} / \epsilon_K$) as well,
which is bigger than its experimental value \cite{Baek:1999jq,Baek:2001kc}.
The other mass insertion
$\de{13}{RR}$ may modify
$B^0$--$\overline{B^0}$ mixing to a sizeable extent.

One can make an observation from Eq.~(\ref{eq:1213}) that
  $\de{12}{RR}$ and $\de{13}{RR}$
will be highly suppressed
provided that neutrino Yukawa couplings have
strong hierarchy and that $[V_L]_{31}$ is vanishingly small.
In fact, the former condition is naturally realized in a scenario
with SO(10) unification \cite{chang}.
In this scenario, the neutrino Yukawa couplings are unified with
the up-type quark Yukawa couplings, thereby resulting in
$y_{\nu_2}^2 / y_{\nu_3}^2 \sim 10^{-4}$.
Since $y_{\nu_2}^2 / y_{\nu_3}^2$ is much smaller than
$m_{\nu_2} / m_{\nu_3}$, 
eigenvalues of $M_N$ should be split accordingly
for the seesaw formula to yield correct neutrino mass spectrum.
Then the RG induced 
$\de{12}{RR}$ and $\de{13}{RR}$ will be small, and SUSY
contributions to Re ($\epsilon^{'} / \epsilon_K$), and
$K^0$--$\overline{K^0}$ and $B^0$--$\overline{B^0}$ 
mixings can be safely neglected.
We will consider the correlation between
$S_{\phi K}$ and Re ($\epsilon^{'}/\epsilon_K$) in such a case, relegating
the more general case with sizable $\de{12}{RR}$ and $\de{13}{RR}$ insertions
to a future study \cite{work}.  

This paper is organized as follows. In Sec.~II, we give
basic formulae for the SUSY contributions for $\Delta B = 1$ effective
Hamiltonian which are relevant in the mSUGRA boundary conditions, 
and discuss how to include hadronic uncertainties in $S_{\phi K}$. 
In Sec.~III, we give the relevant information on $\Delta S= 1$ effective
Hamiltonian and hadronic uncertainties related with 
Re ($\epsilon^{'}/\epsilon_K$). The numerical analysis is given in Sec.~IV,
and the results are summarized in Sec.~V. 

\section{CP asymmetry in $B_d \rightarrow \phi K_S$}

Let us start with  $B_d \rightarrow \phi K_S$ decay 
by recapitulating the effective Hamiltonian for $\Delta B = 1$
relevant to  $B_d \rightarrow \phi K_S$ ($S_{\phi K}$). 
With the operator basis in Eqs.~(9) of Ref.~\cite{Kane:2002sp}, it is given by  
\begin{equation}
\label{eq:hdb1}
H^{\Delta B = 1}_{\rm eff} = \frac{G_F}{\sqrt{2}}\sum_{p=u,c} 
\lambda_p^{(s)} \left[ C_1 O_1^p + C_2 O_2^p + \sum_{i=3}^{10} 
C_i ( \mu ) O_i ( \mu ) + C_{7 \gamma} O_{7\gamma} + C_{8g} O_{8 g} \right]
+ {\rm H.c.}
\end{equation}
One also has tilded operators $\widetilde{O}_{i=3,\ldots,10,7\gamma, 8g}$, which 
are obtained from $O_i$'s 
by making chirality flip $L\leftrightarrow R$. 
Then the Wilson coefficients for the tilded (chromo)magnetic operators,
up to the second order of mass insertion approximation, [ Fig.~1 (a) ], 
read 
\begin{equation}
  \label{eq:c7tc8t}
  \begin{aligned}
  - \lambda_t \frac{G_F}{\sqrt{2}} \widetilde{C}_{7\gamma} &=
  \frac{\pi \alpha_s}{\widetilde{m}^2} \left[
    - \frac{4}{9} M_4(x) \de{23}{RR} 
    + \frac{4}{9} \frac{m_{\tilde{g}}}{m_b} 
       M_2(x) \de{23}{RR}\de{33}{RL}
    \right] ,
    \\
  - \lambda_t \frac{G_F}{\sqrt{2}} \widetilde{C}_{8g} &
  = \frac{\pi \alpha_s}{\widetilde{m}^2} \biggl[
    \left(
      \frac{3}{2} M_3(x) - \frac{1}{6} M_4(x)
      \right) \de{23}{RR} \\
      &
      - \frac{m_{\tilde{g}}}{m_b}
      \left(
            - \frac{1}{6} M_2(x) + \frac{3}{2} M_1(x)
        \right)  
        \de{23}{RR} \de{33}{RL}
    \biggr] .
  \end{aligned}
\end{equation}
The loop functions for single and  double mass insertions can be found 
in Refs.~\cite{Gabbiani:1996hi} and \cite{Baek:2001kc}, respectively.
We  also include the contributions  from $\widetilde{C}_{3,\ldots,6}$. 
We have ignored the terms depending on $\de{23}{RL}$ in 
the tilded Wilson coefficients, and  
$\de{23}{LL}$ and $\de{23}{LR}$ in the untilded Wilson coefficients, 
because they are all small within mSUGRA scenarios \cite{Hagelin:1992tc}.

We calculate $S_{\phi K}$ using the QCD factorization in the BBNS 
approach \cite{bbns}. 
There are theoretical uncertainties from the divergent integral in the 
hard-scattering ($H$) and the weak annihilation ($A$) contributions 
such as  $\int_0^1 dy / y$. We adopt the suggestion by BBNS as follows
 \cite{Beneke:2001ev}: 
\[
\int_0^1 dy / y \rightarrow 
( 1 + \varrho_{H,A} e^{i \varphi_{H,A}} ) \log ( m_B / \lambda_h )
\]
with  
\begin{equation}
  0 \le \varrho_H,\ \varrho_A \le 1, \quad
  0 \le \varphi_H,\ \varphi_A < 2 \pi .
\end{equation}
This prescription is an intrinsic limitation of the BBNS approach, and 
the associated uncertainties cannot be reduced at the moment.
It turns out that these uncertainties are not very large if squarks and 
gluinos are relatively heavy, 
$350 {\rm GeV} \lesssim \tilde{m}, m_{\tilde{g}}$, but can be large for 
lighter squarks and gluinos close to the current lower bounds. 
If we assume the gaugino mass unification within SUSY GUT's, the LEP bound
on chargino ($m_{\chi^+} > 94 $ GeV)
implies that $m_{\tilde{g}} > 400$ GeV. 
In the numerical analysis, we  use $m_{\tilde{g}} = \tilde{m} =  500$ GeV, and
the hadronic uncertainties become smaller.

Finally, we will impose the usual bounds that any new physics involving $b 
\rightarrow s$ transitions should
satisfy \cite{Stocchi:2000ps}:
\begin{equation}
\begin{aligned}
  2.0 \times 10^{-4} & \le B (B\rightarrow X_s \gamma)
  \le 4.5 \times 10^{-4}, \\
  & \Delta M_s \ge 14.9 \ \mathrm{ps}^{-1} .
\end{aligned}
\end{equation}
  We should comment on constraints from
  $b \rightarrow s\,\ell^+ \ell^-$,
  another important ingredient.
  The branching ratio of the exclusive decay mode
  $B \rightarrow K \ell^+ \ell^-$ is \cite{Eidelman:2004wy}
  \begin{equation}
    B (B \rightarrow K \ell^+ \ell^-) = ( 5.4 \pm 0.8 ) \times 10^{-7} ,
  \end{equation}
  which is consistent with the SM prediction \cite{Ali:1999mm}.
  A model independent analysis in Ref.~\cite{Ali:1999mm} has shown that
  the region on the plane of $(C_9^\text{NP}, C_{10}^\text{NP})$
  allowed by this decay at the 90\% C.L.
  is an annulus with radius $\sim 7$
  and thickness $\sim 5$ including the origin,
  where $C_9^\text{NP}$ and $C_{10}^\text{NP}$ are
  new physics contributions to
  the Wilson coefficients
  of the $b \rightarrow s\,\ell^+ \ell^-$ four fermion operators
  under their convention.
  On the other hand,
  for an $O(1)$ value of $|\de{23}{LL}|$ and 
  $m_{\tilde{g}} = \tilde{m} =  500$ GeV,
  a maximal gluino-squark loop contribution
  to $C_9^\text{NP}$ has the size $\sim 0.2$, which is
  much smaller than the extent of the allowed annulus.
  Although the data quoted in the above reference is
  less precise than the present one, the overall feature should
  remain the same.
  Also, incorporation of the inclusive mode does not make much difference.
  From this we can deduce that
  the effect of an $RR$ insertion,
  which we are considering in this work,
  should be equally insignificant.
  As for the effective $RL$ insertion,
  it is much more severely constrained by $B \rightarrow X_s \gamma$
  than $B \rightarrow X_s \ell^+ \ell^-$.
  For these reasons, we do not include the $b \rightarrow s\,\ell^+ \ell^-$
  constraints explicitly in our analysis.

\section{
${\rm Re}~( \epsilon^{'} / \epsilon_K )$}

The $\Delta S = 1$ effective Hamiltonian is given by
\begin{equation}
  H^{\Delta S = 1}_\text{eff} = 
  C_1 O_1 + C_2 O_2 + \sum_{i=3}^{10} C_i O_i
  + C_{7\gamma} O_{7\gamma} + C_{8g} O_{8g} + \mathrm{h.c.} ,
\end{equation}
where the operators $O_{7\gamma}$ and $O_{8g}$ are
the same as in Eq.~(\ref{eq:hdb1}) except the replacements
$(s, b) \rightarrow (d, s)$.
The leading contributions to the Wilson coefficients of the 
(chromo)magnetic operators are provided by the triple mass insertions 
[ Fig.~1 (c) ],  reading as  
\begin{equation}
  \label{eq:c7c8kaon}
  \begin{aligned}
    C_{7\gamma} &=
  \frac{\pi \alpha_s}{\widetilde{m}^2} 
    \frac{4}{9} \frac{m_{\tilde{g}}}{m_s} 
    N_1(x) \de{13}{LL}\de{33}{LR}\de{32}{RR}
    ,
    \\
    C_{8g} &=
  \frac{\pi \alpha_s}{\widetilde{m}^2} 
        \frac{m_{\tilde{g}}}{m_s} 
          \left[
        \frac{1}{6} N_1(x) 
         + \frac{3}{2} N_2(x)
          \right] 
         \de{13}{LL} \de{33}{LR} \de{32}{RR} .
  \end{aligned}
\end{equation}
The loop functions $N_1(x)$ and $N_2(x)$ are available in 
Ref.~\cite{Hisano:2004tf}.  In these expressions, 
we have omitted double and triple insertion terms from other mass insertion
parameters as they are small compared to the terms we kept above. 
As in $C_{7\gamma}$ and $C_{8g}$,
we ignore the $\de{12}{LL}$ contributions in $C_{3,\ldots,6}$.
The importance of the single LR insertion and the double insertion 
contributions to 
Re ($\epsilon^{'} / \epsilon_K$) was pointed out in 
Refs.~\cite{Masiero:1999ub} and \cite{Baek:1999jq,Baek:2001kc}, 
respectively, in the general MSSM frameworks without relying on the 
flavor universal boundary conditions.
In the following, we show that the triple mass insertion can give an 
important contribution to Re ($\epsilon^{'} / \epsilon_K$).

After accomplishing the computation of the relevant SUSY contributions 
in the effective hamiltonian responsible for 
Re ($\epsilon^{'} / \epsilon_K$),  we now turn to the delicate 
issue of the hadronic uncertainties in the evaluation of
Re ($\epsilon^{'} / \epsilon_K$). For definiteness we will  follow the 
treatment provided in  Refs.~\cite{Buras:2003zz,Buras:1999da}.
In the SM contribution, 
the main uncertainties reside in the evaluation of the $B$ parameters  
$B_6^{(1/2)}$, $B_8^{(3/2)}$ and in the estimate of the strange 
quark mass $m_s$.
We define the non-perturbative parameters $R_6$ and $R_8$ as
\begin{equation}
  R_6 \equiv B_6^{(1/2)}
  \left[ \frac{121\ \text{MeV}}{m_s(m_c) + m_d(m_c)} \right ]^2 ,
  \quad
  R_8 \equiv B_8^{(3/2)}
  \left[ \frac{121\ \text{MeV}}{m_s(m_c) + m_d(m_c)} \right ]^2 .
\end{equation}
The value of $B_8^{(3/2)}$ is rather well estimated both from
lattice QCD \cite{Becirevic:2002mm}
and from analytic non-perturbative approaches 
\cite{B8analytic}.
In what follows, we employ the range of $R_8$,
\begin{equation}
  R_8 = 1.0 \pm 0.2 .
\end{equation}
On the other hand, the situation of $B_6^{(1/2)}$ is
very unclear, and there exist results from different approaches
ranging within a factor of 2.2. For instance,  the large-$N_c$ limit 
predicts  $R_6 = 1$. It is difficult to attribute a reliable uncertainty
to such estimates. Taking $R_6 = R_8 = 1.0$ results in the SM prediction 
of Re ($\epsilon^{'} / \epsilon_K$), which is 
smaller than the observed value by about $3 \times 10^{-4}$.
According to Ref.~\cite{Buras:2003zz}, the best fit from
Re ($\epsilon^{'} / \epsilon_K$) yields  $R_6 = 1.23$.
However, let us emphasize that this is the case within the SM, and we 
cannot rely on
a SM fit here in the presence of new physics affecting kaon decays.
As a representative value, we use
\begin{equation}
  R_6 = 1.0 \pm 0.2 .
\end{equation}
In any event, our conclusion does not depend significantly on the
precise value of $R_6$ or $R_8$ because
Re ($\epsilon^{'} / \epsilon_K$) is dominated by chromomagnetic
contribution in most of the parameter space.

As for the strange quark mass, we make use of the range:
\begin{equation}
  m_s (m_c) = 115 \pm 20\ \text{MeV}.
\end{equation}
For completeness, we also specify our choice of the isospin breaking 
parameter $\Omega_\text{IB}$. We use the value
\cite{Cirigliano:2003nn}
\begin{equation}
  \Omega_\text{IB} = 0.06.
\end{equation}
The uncertainty of $\Omega_\text{IB} \sim 0.08$ could shift
Re ($\epsilon^{'} / \epsilon_K$) by about $0.8 \times 10^{-4}$,
and neglecting it does not affect our conclusion.

It is time to come to the main uncertainty in the SUSY contribution to
Re ($\epsilon^{'} / \epsilon_K$). This is related to the evaluation of the 
matrix element:
\begin{equation}
  \langle Q_g^- \rangle _0 = \sqrt{\frac{3}{2}} \frac{11}{16 \pi^2} \,
  \frac{\langle \overline{q} q \rangle}{F_\pi^3} \, m_\pi^3 \, B_G ,
\end{equation}
where
\begin{equation}
  Q_g^- = \frac{1}{4 m_s} \left(
    \widetilde{O}_{8g} - O_{8g}
  \right) .
\end{equation}
The uncertainty in the above evaluation is encoded in the value of 
the parameter $B_G$. The
result of Ref.~\cite{Bertolini:1994qk} 
corresponds to $B_G$ = 1. Unfortunately lattice computations are still 
unable to come up with a reliable estimate of the matrix element 
$\langle Q_g^- \rangle _0$;  in fact, even the sign of this parameter
is not certain yet,
although the above reference estimated it to be positive.
If we assume the opposite sign of $B_G$ with the same magnitude,
the SUSY contribution 
flips its sign. In spite of all this and even allowing for an uncertainty 
of a factor 4 in the estimate of $|B_G| = \text{1--4}$, we will show that   
the constraint on $( \delta_{23}^d )_{RR}$ from Re ($\epsilon^{'} / 
\epsilon_K$) still remains meaningful.

\section{Numerical analyses}

Now it is straightforward to calculate Re ($\epsilon^{'} / \epsilon_K$)  
and $S_{\phi K}$ from $s \rightarrow d g$ and $b \rightarrow s g$,
when the $\de{23}{RR}$ is the main new physics contribution beyond 
the SM contributions. 
Here we assume that SUSY contributions from $\de{12}{RR}$ and
$\de{13}{RR}$ are negligible as is the case
for hierarchical neutrino Yukawa couplings and vanishing $[V_L]_{31}$.
If we relax this assumption, we should regard $\gamma$ as a free
variable that can be varied within a range compatible with
other data such as $B \rightarrow X_d \gamma$.
Then, the calculated value of Re ($\epsilon^{'} / \epsilon_K$)
will change by a fraction of its SM prediction.
Nevertheless, the qualitative feature remains true that
the size of Re ($\epsilon^{'} / \epsilon_K$) bounds
the deviation of $S_{\phi K}$ from $S_{\phi K}^{\rm SM}$.

We would like to point out the main result of this work 
in a simple way, before we give a detailed numerical analysis. 
If the $( \delta^d_{23} )_{RR}$ mixing is the dominant new physics 
contribution to $B_d \rightarrow \phi K_S$, we find the following from 
Eqs.~(\ref{eq:c7c8kaon}) and (\ref{eq:c7tc8t}) in the previous sections :
\begin{eqnarray}
{\rm Re} (\epsilon^{'}/\epsilon_K) : 
C_{8g}^{\rm SUSY} ( \Delta S = 1 ) & \propto & 
f_{1} (x)~\de{13}{LL} \de{33}{LR} \de{32}{RR} ,   
\\
S_{\phi K} : \widetilde{C}_{8g}^{\rm SUSY} ( \Delta B = 1 ) & \propto & 
f_{2} (x)~\de{23}{RR} + f_{3} (x)~ { m_{\tilde{g}} \over m_b}~
\de{33}{RL} \de{23}{RR} ,
\end{eqnarray}
where $f_{i=1,2,3} (x)$ are the loop functions obtained in the previous
sections. 
Now, if the SUSY contribution saturates Re ($\epsilon^{'}/\epsilon_K$), 
then  it is well known that  one has to satisfy 
\[
| \de{13}{LL} \de{33}{LR} \de{32}{RR} | \lesssim 10^{-5}
\]
with an $O(1)$ phase \cite{Gabbiani:1996hi}.
Since the RG evolution generates $\de{13}{LL} \sim \lambda^3$ within 
mSUGRA scenario, we can  derive the following upper bound:
\begin{equation}
\label{eq:constraint}
| \de{33}{LR} \de{32}{RR} | \lesssim 10^{-3}.   
\end{equation}
Note that this combination enters the calculation
of $S_{\phi K}$ and $B\rightarrow X_s \gamma$ 
through $C_{8g(7\gamma)} ( \Delta B = 1)$ along with $\de{32}{RR}$.
For a small $\mu\tan\beta$ (corresponding to a small $\de{33}{RL}$), 
one can have larger $\de{32}{RR}$, which is constrained by the lower bound 
on $\Delta M_s$  and the $B\rightarrow X_s \gamma$ branching ratio. 
For a large $\mu \tan\beta$ (corresponding to a large $\de{33}{RL}$), 
$\de{32}{RR}$ should be smaller in order to satisfy (\ref{eq:constraint}).
In either case, we can expect that the deviation in $S_{\phi K}$ cannot 
be that large for such $\de{32}{RR}$  satisfying the 
Re ($\epsilon^{'}/\epsilon_K$) constraint, (\ref{eq:constraint}).

Having described the qualitative features of our main points, we now
provide the detailed  analysis including theoretical uncertainties 
in $S_{\phi K}$ and Re ($\epsilon^{'}/\epsilon_K$) as summarized in 
Secs.~II and III.
We use the parameterization, $\de{23}{RR} \equiv r \, e^{i \phi}$. 
We fix the modulus $r$ at a maximal 
value compatible with $B (B\rightarrow X_s \gamma)$, 
and vary its phase $\phi$ from 0 to $2 \pi$.
For $\mu \tan \beta = 1$ TeV, we use $r = 1$ which is dictated by 
requiring the validity of the mass insertion approximation.
For $\mu \tan \beta = 5$ TeV, we set $r = 0.33$,
which is defined by the upper bound on $B (B\rightarrow X_s \gamma)$.
Some values of $\phi$ result in $\Delta M_s$
smaller than the lower bound \cite{Kane:2002sp}, and they are discarded.
For each value of $\phi$ consistent with the $\Delta M_s$ constraint,
we plot a point in  the plane 
Re ($\epsilon^{'} / \epsilon_K$) -- $S_{\phi K}$, 
following the procedures in Refs.~\cite{Baek:2001kc}
and \cite{Kane:2002sp,Ciuchini:2002uv}, respectively.
 In Figs.~2 (a) and (b), we show  the  plots  
for   $\mu \tan \beta = 1$ and $5$ TeV, respectively, 
with $\tilde{m} = m_{\tilde{g}} = 500$ GeV.
The thick vertical error bar shows the current data on
$S_{\phi K}$, 
and the two dashed vertical lines delimit the experimental value
of Re ($\epsilon^{'} / \epsilon_K$) \cite{Eidelman:2004wy},
\begin{equation}
  \mathrm{Re}\ (\epsilon^{'} / \epsilon_K)
  = ( 16.7 \pm 2.6 ) \times 10^{-4} .
\end{equation}
The full black box shows our estimates of $S_{\phi K}$ and
Re ($\epsilon^{'} / \epsilon_K$) within the SM.  
Its width and height are the uncertainties
in Re ($\epsilon^{'} / \epsilon_K$) and $S_{\phi K}$, respectively.
In Fig.~2~(a), we show a curve for $\mu \tan \beta = 1$ TeV,
and $r = 1$ which may be regarded to define the boundary of a region 
of the two observables generically  predicted in this scenario.
For this curve, we fix $B_G = 2$ and use the central values of
$R_6$, $R_8$, and $m_s$, as given in Sec.~III.
As mentioned previously, disconnected parts of the curve
are excluded by the $\Delta M_s$ constraint.
If we turn on hadronic uncertainties, this curve gets
broadened into the gray region around it.
We estimate the uncertainty of an observable by taking
its maximum and minimum values reached while varying
the relevant input parameters in the ranges quoted in the previous two sections.
For Re ($\epsilon^{'} / \epsilon_K$), they are
$R_6$, $R_8$, $m_s$, and $B_G$,
and for $S_{\phi K}$, they are $\varrho_H$, $\varphi_H$,
$\varrho_A$, and $\varphi_A$.
Each of the uncertainties in Re ($\epsilon^{'} / \epsilon_K$)
and $S_{\phi K}$ is displayed by the horizontal or vertical
error bars at five selected points.
The gray region is drawn by varying all the eight parameters
simultaneously.
It therefore covers wider space than is obtained by
quadrature addition.
Suppose that the sign of $B_G$ is negative.
The resulting curve and the region around it can easily be
guessed by taking the mirror image of the present one
around the vertical axis passing through the SM point.
Even then the Re ($\epsilon^{'} / \epsilon_K$) data
gives a strong constraint on the possible value of $S_{\phi K}$.

We repeat the same exercise for $\mu \tan \beta = 5$ TeV in
Fig.~2~(b). We find that $S_{\phi K} > 0.25$ if the data on 
Re ($\epsilon^{'} / \epsilon_K$) is imposed. 

Note that the constraint from Re ($\epsilon^{'} / \epsilon_K$) is
comparable to that from the strange quark CEDM.
In particular the positive (negative)  $S_{\phi K}$ is correlated 
with the positive (negative) Re ($\epsilon^{'} / \epsilon_K$) 
within minimal SUGRA boundary conditions. Therefore, the old Belle
data with the negative $S_{\phi K}$ implies a negative
Re ($\epsilon^{'} / \epsilon_K$) in the $RR$ dominance scenario such as 
SUSY GUT models with right-handed neutrinos, which is clearly excluded 
by the data Re $(\epsilon^{'}/\epsilon_K) = (16.7 \pm 2.6 ) \times 10^{-4}$.  
If the old Belle data were still valid, then the $RR$ dominance scenario
should have been discarded. 
Our results provide a meaningful correlation between $S_{\phi K}$ and
Re ($\epsilon^{'} / \epsilon_K$) despite large hadronic uncertainties
in both quantities. This is independent of the strange  quark CEDM 
constraint, and probably has less theoretical uncertainties.

One may wonder why we are not considering triple mass insertion
on a squark line in the box diagram for $\epsilon_K$.
The size of the effective insertion,
\begin{equation}
  \de{12}{LR}^\mathrm{eff} \equiv \de{13}{LL} \de{33}{LR} \de{32}{RR} ,
\end{equation}
is always smaller than $2 \times 10^{-4}$ due to 
$B (B\rightarrow X_s \gamma)$.
If we require that the SUSY contribution to
Re ($\epsilon^{'} / \epsilon_K$) be smaller than its experimental value,
we get
\begin{equation}
  \left| \mathrm{Im} \de{12}{LR}^\mathrm{eff} \right| \lesssim 10^{-5} ,
\end{equation}
which implies that
\begin{equation}
  \sqrt{ \left| \mathrm{Im} [ \de{12}{LR}^\mathrm{eff} ]^2 \right| }
  \lesssim 6 \times 10^{-5} .
\end{equation}
This limits the SUSY contribution to $\epsilon_K$ below
1/30 of its experimental value.
In view of theoretical uncertainties in predicting $\epsilon_K$,
we may regard $\epsilon_K$ being always safe provided that
Re ($\epsilon^{'} / \epsilon_K$) constraint is satisfied.

Let us add a remark on the $CP$ violation in $B_s$--$\overline{B_s}$
mixing, whose indirect $CP$ asymmetry nearly vanishes in the SM.
The large $RR$ mixing leads to a considerable modification
in the mixing amplitude.
This is evident from the fact
that part of the curves in Figs.~2 was excluded by the
$\Delta M_s$ constraint.
Since this new piece of amplitude has a phase different from
the SM one in general,
the indirect $CP$ asymmetry in $B_s$--$\overline{B_s}$ mixing
can have a value of $O(1)$ according to the phase of $\de{23}{RR}$.
This will show up in the decay channel $B_s \rightarrow J/\psi \phi$
as the time dependent $CP$ asymmetry therein.
For quantitative analyses, see
Refs.~\cite{Kane:2002sp,Harnik:2002vs,Ciuchini:2002uv} for instance.

In SU(5) SUSY GUT,
the left-handed sleptons and the right-handed down-type squarks
are tied to a {\boldmath$\overline{5}$}, and therefore
their flavor changing effects are related to each other.
Off-diagonal elements of the left-handed slepton mass matrix
are given by
\begin{equation}
  \begin{aligned}
    ( 
  m^2_{\tilde{l}} )_{ij} &\simeq - \frac{1}{8 \pi^2} 
  y_{\nu_k}^2 [V_L]_{ki} [V_L^*]_{kj}
  ( 3 m_0^2 + A^2 ) \log \frac{M_*}{M_{N_k}} ,
  \label{eq:m2lij}
  \end{aligned}
\end{equation}
in a way similar to Eq.~(\ref{eq:m2dij}).
In order to estimate lepton flavor violation from this
mass matrix, we should go to the super CKM basis.
If the down-type quark and the charged lepton Yukawa matrices are
the same at the GUT scale as
\begin{equation}
  \label{eq:yukawa-unif}
  Y_d = Y_l^T ,
\end{equation}
the right-handed down-type squark and the charged slepton
mass insertion parameters are unified as well.
Under this assumption, the current upper bound on
the $\tau \rightarrow \mu \gamma$ branching ratio \cite{Aubert:2005ye},
\begin{equation}
  B (\tau \rightarrow \mu \gamma) < 6.8 \times 10^{-8} ,
\end{equation}
translates into the limit \cite{Ciuchini:2003rg},
\begin{equation}
  |\de{23}{RR}| < 0.03 ,
\end{equation}
for $\tan \beta = 10$.
This is roughly 1/10 of the size of $\de{23}{RR}$ we
used for $\mu \tan \beta = 5$ TeV,
and we cannot expect considerable change in $S_{\phi K}$
satisfying the $\tau \rightarrow \mu \gamma$ constraint.
The assumption of Yukawa unification, however,
leads to an incorrect mass relation,
\begin{equation}
  \frac{m_d}{m_s} = \frac{m_e}{m_\mu} ,
\end{equation}
and Eq.~(\ref{eq:yukawa-unif}) should be modified
to account for the mass ratios of the first and second generation fermions.
Even in this case,
a mass insertion involving a third generation
is not much affected, and $\tau \rightarrow \mu \gamma$
remains a strong constraint on $S_{\phi K}$.


In this work, we are considering SUSY GUT with
right-handed neutrinos as an example of
a scenario that gives rise to large $RR$ mixing.
This is why $\tau \rightarrow \mu \gamma$ constrains
$S_{\phi K}$.
Here, we would like to stress again that the interconnection between
Re($\epsilon^{'}/\epsilon_K$) and $S_{\phi K}$,
unlike $\tau \rightarrow \mu \gamma$,
is a common consequence of large $RR$ mixing,
which is not specific to SUSY GUT.
Suppose that there is a large $RR$ mixing but no $LL$ mixing
in the squark sector at the reduced
Planck scale due to a flavor symmetry and that
we do not have a unified gauge group.
Even in this case, one has a strong correlation between
Re($\epsilon^{'}/\epsilon_K$) and $S_{\phi K}$,
while $\tau \rightarrow \mu \gamma$ is unrelated to
$S_{\phi K}$.

If we considered more general scalar masses at $M_*$ with some flavor 
structures, then our results will be changed accordingly. 
The Wilson coefficients for $C_{8g}$'s for both $\Delta B (S) =1$ have
to include other mass insertion parameters such as $\de{23}{LR}$,
$\de{12}{LL}$, $\de{23}{LL}$, etc., which were neglected in Secs.~II and 
III because they are small within the mSUGRA scenarios. Still we should 
make it sure that  the new flavor physics that affects $S_{\phi K}$ 
does not contribute to Re($\epsilon^{'}/\epsilon_K$) too much, and 
this could make a strong constraint on new sources of flavor and CP 
violation despite theoretical uncertainties in
Re($\epsilon^{'}/\epsilon_K$). 

\section{Conclusions}

In summary, we showed that if the $RR$ $b\rightarrow s$ transition is
large with $O(1)$ phase, it can affect not only $S_{\phi K}$ through
double mass insertion and the strange quark CEDM through triple
mass insertion, but it also affects Re ($\epsilon^{'} / \epsilon_K$).
The correlation between the two observables are  strong despite large
hadronic uncertainties in both observables, within mSUGRA boundary 
conditions with flavor universal scalar masses at $M_*$. 
The current data on Re ($\epsilon^{'} / \epsilon_K$) indicates that
$S_{\phi K}$ should be in the range of $0.25$--$1.0$, which is now 
in accord with the present world average of $S_{\phi K}$.
\begin{acknowledgments}
We are grateful to Weon Jong Lee and A.I. Sanda for discussion on the 
Re $( \epsilon^{'} / \epsilon_K )$. 
PK and AM thank Aspen Center for Physics for the hospitality during their 
stay where this work was initiated. 
PK is supported in part by the BK21 Haeksim Program, 
KOSEF through CHEP at Kyungpook National  University, 
and by KOSEF Sundo Grant R02-2003-000-10085-0. 
\end{acknowledgments}

\begin{figure}
\includegraphics[width=8cm]{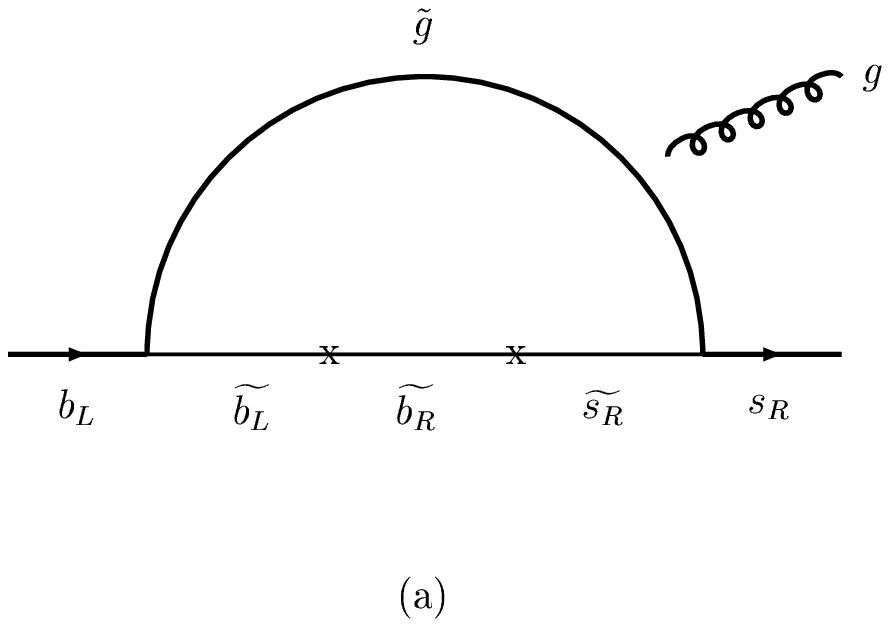}\\
\vspace{1cm}
\includegraphics[width=8cm]{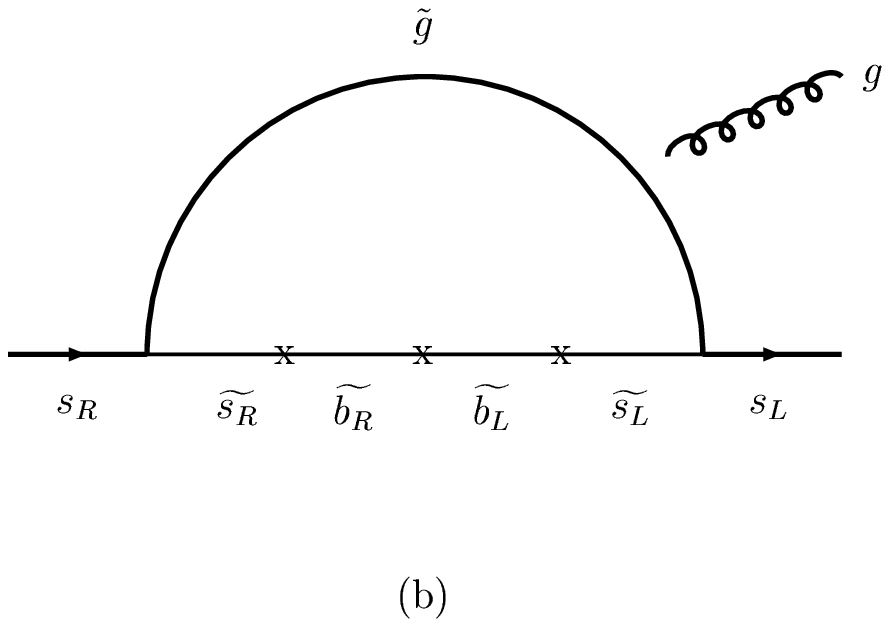}\\
\vspace{1cm}
\includegraphics[width=8cm]{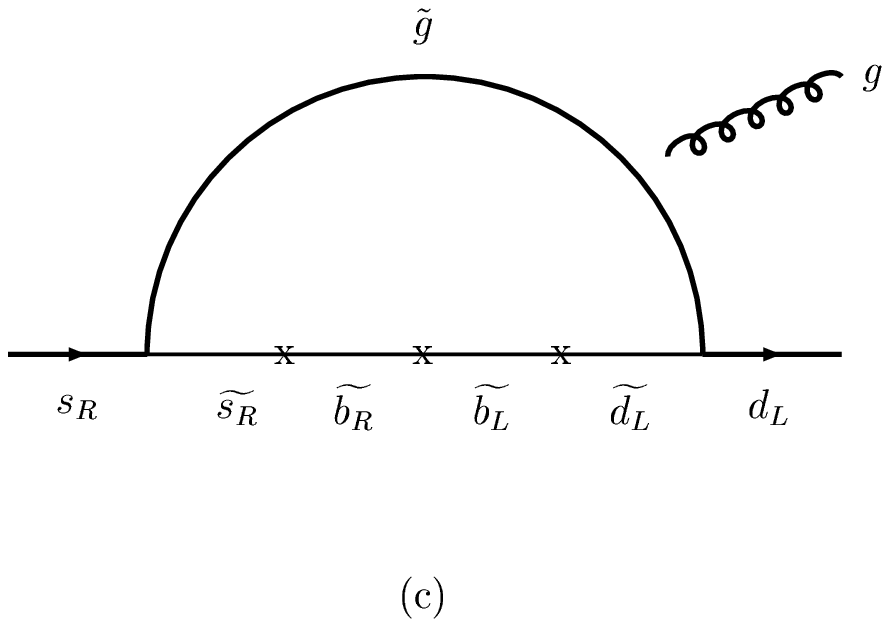}%
\caption{\label{} Feynman diagrams for (a) $b\rightarrow s g$ with double
mass insertions, (b) strange quark CEDM with triple mass insertions, 
and (c) $s\rightarrow d g$ with triple mass insertions,
involving $( \delta^d_{23} )_{RR}$ as the dominant source of
new  CP violating parameter contributing to $S_{\phi K}$ and 
{\rm Re} ($\epsilon^{'} / \epsilon_K$).
}
\end{figure}

\begin{figure}
\subfigure[$\mu \tan \beta = 1$ TeV]%
{\includegraphics[width=8cm]{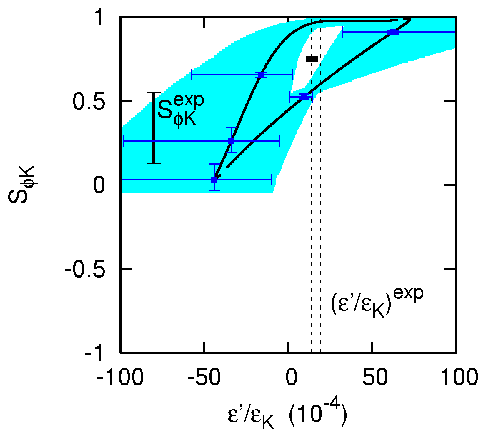}}
\subfigure[$\mu \tan \beta = 5$ TeV]%
{\includegraphics[width=8cm]{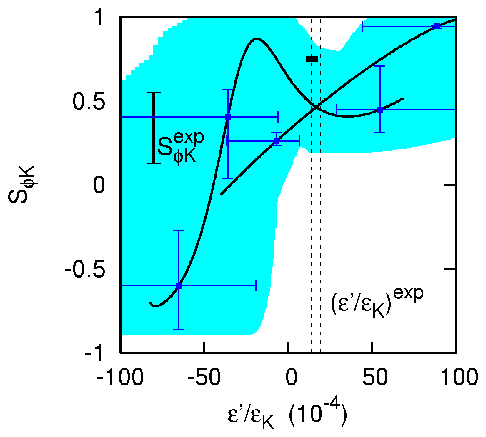}}%
\caption{
$S_{\phi K}$ vs. {\rm Re} ($\epsilon^{'} / \epsilon_K$) for 
(a) $\mu \tan \beta = 1$ TeV and (b) $\mu \tan \beta = 5$ TeV,
with $\tilde{m} = m_{\tilde{g}} = 500$ GeV.
Experimental bounds on {\rm Re} ($\epsilon^{'} / \epsilon_K$)
and $S_{\phi K}$ are depicted by the vertical dashed lines
and the thick vertical error bar, respectively.
Their SM predictions
are marked by the black box, whose extent indicates
their uncertainties.
The black curve does not include hadronic uncertainties, and
the gray region includes them.
The respective uncertainties in {\rm Re} ($\epsilon^{'} / \epsilon_K$)
and $S_{\phi K}$ are shown by the horizontal and
vertical error bars at some selected points.
}
\end{figure}

\end{document}